\documentclass[epj]{svjour}
\usepackage{graphics}

\begin{document}

\title{Breakup of loosely bound nuclei as indirect method in nuclear
astrophysics: $^{8}$B, $^{9}$C, $^{23}$Al }
\author{L. Trache\inst{1} \and F. Carstoiu\inst{2} \and C.A. Gagliardi%
\inst{1} \and R.E. Tribble \inst{1} }
\institute{Cyclotron Institute, Texas A\&M University, College Station, TX 77843-3366,
USA \and National Institute of Physics and Nuclear Engineering H. Hulubei,
Bucharest, Romania }
\date{Received: date / Revised version: date
} 


\abstract{
We discuss the use of one-nucleon breakup reactions of loosely bound nuclei
at intermediate energies as an indirect method in nuclear astrophysics.
These are peripheral processes, therefore we can extract asymptotic
normalization coefficients (ANC) from which reaction rates of astrophysical
interest can be inferred. To show the usefulness of the method, three
different cases are discussed. In the first, existing experimental data for
the breakup of $^{8}$B at energies from 30 to 1000 MeV/u and of $^{9}$C at
285 MeV/u on light through heavy targets are analyzed. Glauber model
calculations in the eikonal approximation and in the optical limit using
different effective interactions give consistent, though slightly different
results, showing the limits of the precision of the method. The results lead
to the astrophysical factor $S_{17}(0)=18.7\pm 1.9$ eVb for the key reaction
for solar neutrino production $^{7}$Be(p,$\gamma $)$^{8}$B. It is consistent
with the values from other indirect methods and most direct measurements,
but one. Breakup reactions can be measured with radioactive beams as weak as
a few particles per second, and therefore can be used for cases where no
direct measurements or other indirect methods for nuclear astrophysics can
be applied. We discuss a proposed use of the breakup of the proton drip line
nucleus $^{23}$Al to obtain spectroscopic information and the stellar
reaction rate for $^{22}$Mg(p,$\gamma $)$^{23}$Al.
} 
\maketitle

\section{Introduction}

\label{intro} Radiative proton capture reactions are important in nuclear
astrophysics, and a large number of reaction chains were found to be needed
in nucleosynthesis calculations for static or explosive hydrogen burning
scenarios (see e.g. \cite{bahcall,wies89}). This means that more data
involving proton capture on unstable nuclei are necessary. In some cases
direct experiments are possible, but in many more they are impossible with
the present techniques and even with those of the foreseeable future. We
have to rely on indirect methods instead. In this presentation we discuss
such an indirect method, and we shall concentrate on three particular cases,
first to demonstrate the feasibility, then to show the strengths of the
method and its limits.

Part of the results discussed were published before, when we originally 
proposed to
extract astrophysical $S$-factors from one-nucleon-removal (or breakup)
reactions of loosely bound nuclei at intermediate energies or later \cite%
{tra01,tra02,tra04}. In the present paper, first we use the well studied 
case of $^{8}$B breakup as a benchmark 
to demonstrate the usefulness of the method and show the possibilities of
the Glauber reaction model used. We show that existing experimental data at
energies between 30 and 1000 MeV/nucleon \cite%
{nego,blank,enders,warner,cortina} on a range of light and heavy targets
translate into consistent values of the ANC, which is then used to determine
the astrophysical factor $S_{17}$ (which gives the reaction rate for the $%
^{7}$Be(p,$\gamma $)$^{8}$B reaction of crucial importance for the solar
neutrino question). We show that the precision of the method is limited to
about 10\% by our ability to compute absolute cross sections. Second, we use
the same technique for $^{9}$C breakup data at 285 MeV/nucleon \cite{blank}
to determine $S_{18}$ (which gives the rate for the $^{8}$B(p,$\gamma $)$%
^{9} $C reaction of importance for explosive hydrogen burning) with reasonable
accuracy. For a third case, a proposed experiment for the breakup of $^{23}$%
Al is discussed to show that the method is particularly well adapted to rare
isotope beams produced using fragmentation. Spectroscopic information is
sought in this case. In particular we seek to determine the spin and parity 
of the ground state of the dripline nucleus 
$^{23}$Al and the ANC, which will be then used to calculate the 
reaction rates for $^{22}$Mg(p,$\gamma $)$^{23}$Al. 
The last part of the present paper will concentrate on this latter case, 
which has not been discussed before.

\section{The reaction model}

\label{sec:2}

The method is based on data showing that the structure of halo nuclei is
dominated by one or two nucleons orbiting a core \cite{tanihata96,hansen01}.
Consequently, we use the fact that the breakup of halo or loosely bound
nuclei is essentially a peripheral process, and therefore, the breakup
cross-sections can give information about the wave function of the last
nucleon at large distances from the core. More precisely, asymptotic
normalization coefficients (ANCs) can be determined. Then, these ANCs are
sufficient to determine the astrophysical $S$-factors for radiative proton
capture reactions. We show that there exists a favorable kinematical window
in which breakup reactions are highly peripheral and are dominated by the
external part of the wave function and, therefore, the ANC is the better
quantity to be extracted. The approach offers an alternative and
complementary technique to extracting ANCs from transfer reactions \cite%
{mukh01}.

In the breakup of loosely bound nuclei at intermediate energies, a nucleus $%
B=(Ap)$, where $B$ is a bound state of the core $A$ and the nucleon $p$, is
produced by fragmentation from a primary beam, separated and then used to
bombard a secondary target. In measurements, the core $A$ is detected,
measuring its parallel and transverse momenta and eventually the gamma-rays
emitted from its deexcitation. Spectroscopic information can be extracted
from these experiments, such as the orbital momentum of the relative motion
of the nucleon and the contribution of different core states, typically
comparing the measured momentum distributions with those calculated with
Glauber models. The integrated cross sections can be used to extract
absolute spectroscopic factors \cite{hansen01} or the ANC \cite{tra01}. The
latter approach has the advantage that it is independent of the geometry of
the proton binding potential. We note that the ANC $C_{Ap}^{B}$ for the
nuclear system $A+p\leftrightarrow B$ specifies the amplitude of the tail of
the overlap function of the bound state $B$ in the two-body channel $(A\,p)$
(see, for example \cite{mukh01} and references therein). Fortunately, this
ANC is all we need to determine the astrophysical $S$-factor for the
radiative proton capture reaction $A(p,\gamma )B$ which is a highly
peripheral process. Details about the reaction model are published elsewhere 
\cite{tra04,sauvan}.

\section{Three particular cases}

\label{sec:3}

\subsection{Breakup of $^{8}$B to determine the S$_{17}$ astrophysical factor}

The calculations presented in \cite{tra01} have been extended and refined.
The Coulomb part of the dissociation cross section was refined by including
the final state interaction into calculations and new data on the breakup of 
$^{8}$B are analyzed \cite{enders,warner,cortina}. Also a new set of
calculations for the breakup of $^{8}$B were made using five sets of
different effective NN interactions. We describe the breakup of $^{8}$B (and 
in the next subsection of $^{9}$C) in terms of an extended Glauber model. 
The loosely bound $^{8}$B ($%
^{9}$C) nucleus is moving on a straight line trajectory and the proton and
the $^{7}$Be ($^{8}$B) core making it, interact independently with the
target. The breakup cross sections depend on the proton-target and
core-target interactions and on the relative $p$-core motion. The wave
function of the ground state of $^{8}$B ($^{9}$C) is a mixture of $1p_{3/2}$
and $1p_{1/2}$ orbitals, around a $^{7}$Be ($^{8}$B) core. The total ANC $%
C_{tot}^{2}=C_{p_{3/2}}^{2}+C_{p_{1/2}}^{2}$ can be extracted from the
measured breakup cross sections.

\begin{figure}
\resizebox{0.45\textwidth}{!}{ \includegraphics{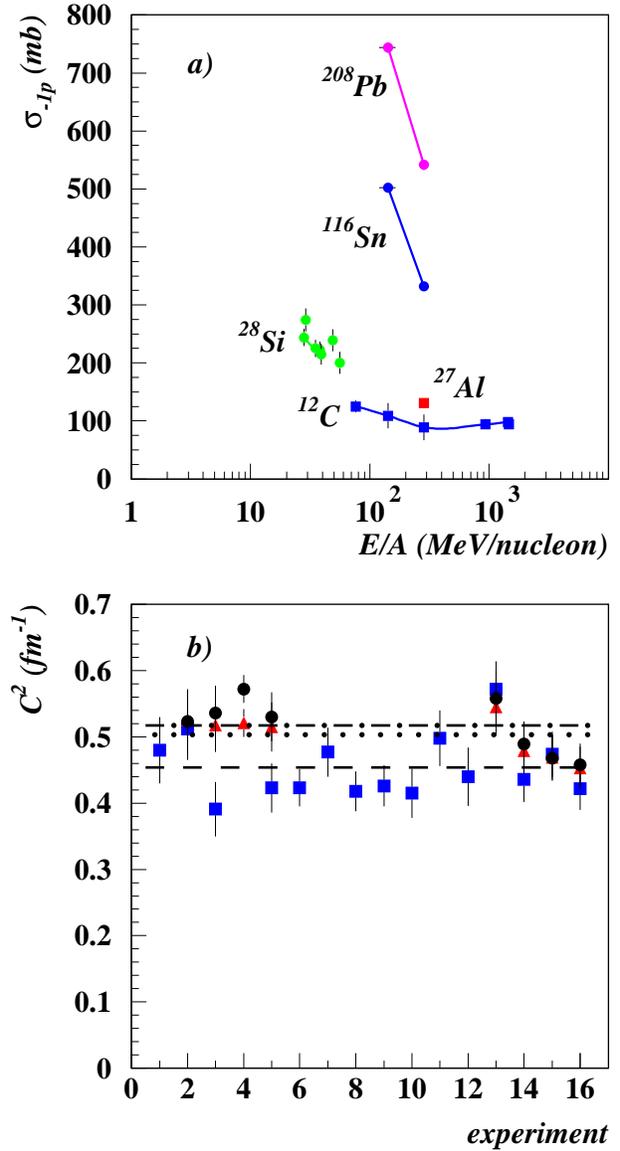}}
\caption{{\protect\footnotesize a) The one-proton-removal cross sections 
on C, Al, Sn and Pb targets, depending on energy. b) The ANCs determined from the 
breakup of $^8$B at 28-1000 MeV/nucleon using the data above and various effective
interactions: JLM (squares), "standard" (circles) and "Ray" (triangles). 
The dashed, dotted and dash-dotted lines are the averages of the three 
interactions above, in that order.}}
\label{fig1}
\end{figure}

The calculations reproduce well all the measured parallel and transverse
momentum distributions measured so far, on light or  
heavy targets, giving us confidence in the Glauber
model used. We show that the reaction is peripheral in various degrees, 
depending on the energy and target used. The $^{8}$B ANC is extracted 
from existing breakup data at
energies between 30-1000 MeV/nucleon and on different targets ranging from C
to Pb \cite{nego,blank,enders,warner,cortina}. Two approaches were used. The
first is a potential approach. To obtain the folded potentials needed in the 
$S$-matrix calculations we used the JLM effective nucleon-nucleon
interaction \cite{JLM}, using the procedure and the renormalizations 
of Ref. \cite{tra04}. We applied this technique for energies below 285
MeV/nucleon only and on all targets. In a second approach, the Glauber model
in the optical limit was used. The breakup process is treated as multiple
elementary interactions between partners' nucleons, and the cross sections
and the complex scattering amplitudes are taken from the literature.
Calculations were done using different ranges for the elementary
interactions: zero range, 1.5 fm ("standard"), 2.5 fm and individual ranges
for each NN component ("Ray") \cite{ray}. No new parameters were adjusted.
The contribution of the $^{7}$Be core excitation was calculated for each target 
and at each energy using the
data from an experiment which disentangle it \cite{cortina}, and corrected 
for in all cases. For details on the
procedure see \cite{tra04}. In Figure \ref{fig1} we show that from the widely varying
breakup cross sections (upper panel) on all targets and at so different energies,
we extract ANCs which are consistent with a constant value (lower panel).
However, we see that a certain dependence on the NN interaction used exists,
which point to the limitations of our present knowledge of the effective 
nucleon-nucleon interactions.

If we take the unweighted average of all 31 determinations we find an ANC $%
C_{tot}^{2}(JLM)=0.483\pm 0.050$ fm$^{-1}$ (Fig. \ref{fig1}). The value is in
agreement with that determined using the ($^{7}$Be,$^{8}$B) proton
transfer reactions at 12 MeV/u \cite{azhari,tabacaru05}. The two values
agree well, in spite of the differences in the energy ranges and in the
reaction mechanisms involved. The ANC extracted leads to the astrophysical
factor $S_{17}(0)=18.7\pm 1.9$ eV$\cdot $ b for the key reaction for solar
neutrino production $^{7}$Be($p$,$\gamma $)$^{8}$B. The uncertainties quoted
are only the standard deviation of the individual values around the average,
involving therefore the experimental and theoretical uncertainties. This
10\% error bar is probably a good measure of the precision we can claim from
the method at this point in time, due essentially to the uncertainties in
the cross section calculations. The $S_{17}$(0) value we extract is also in 
agreement with those extracted from indirect methods and with most of the 
direct determinations (see the discussions in \cite{schumann,davids03,hammache},
but one which stands out in its claim of a larger value and very small error 
\cite{junghans}. There are currently many evaluations of existing or new data 
and variations occur in the central values and uncertainties of the 
determinations. It is difficult to quote all of them and is not our 
intention to do so here. However, we notice that our average value of 
$S_{17}(0)$ 
is very close to the "low" values obtained from Coulomb dissociation data 
and some direct data $S_{17}(0)=18.6\pm 0.4$(exp)$\pm 1.1$(syst) eV b 
\cite{davids03}. It is also in reasonable agreement 
with the average value obtained by Cyburt et al. \cite{cyburt-prc70} 
$S_{17}(0)=20.8 \pm 0.6$(stat)$\pm 1.0$(syst) eV b using all radiative capture data in 
the assumption they are completely independent. The difference between 
our value and the value obtained from the direct measurement of Junghans et al. 
\cite{junghans} $S_{17}(0)=22.1 \pm 0.6$(stat) $\pm 0.6$(theor) 
eV b still exists 
and is only relevant if the small uncertainty of the latter is true, given 
the fact that it involves extrapolation. It would, of course, be interesting 
to understand why the results differ. The difficulties encountered by the 
direct methods, both experimental (very small cross sections, difficult 
targets, etc...) and theoretical (extrapolations), are known. One important 
factor in any indirect determination of the astrophysical S-factor is that 
of the accuracy of the theoretical calculations involved. Much effort is 
done currently, e.g., to investigate the accuracy of the absolute values of 
the calculations used in the analysis of the Coulomb dissociation experiments 
\cite{esbensen-prl94,bertulani-prl94}. We did our part above, using different 
NN-interactions. 
Our central value is about 1$\sigma$ lower than the average central value obtained 
by Cyburt et al. \cite{cyburt-prc70} in a recent analysis that uses 
all of the best available capture data, under the assumption that they are 
independent. Including the uncertainty quoted by Cyburt et al. our results 
are consistent at the 1s level. 

\subsection{Breakup of $^{9}$C to determine S$_{18}$}

The same procedures have been applied for $^{9}$C to determine the
astrophysical $S_{18}$ factor for the reaction $^{8}$B($p$,$\gamma $)${}^{9}$%
C. The reaction is important in hot $pp$-chains as it can provide a starting
point for an alternative path across the $A=8$ mass gap \cite{wies89}. The
ANC for $^{9}\mathrm{C}\rightarrow ^{8}\mathrm{B}+p$, has been determined
using existing experimental data for the breakup of $^{9}$C projectiles at
285 MeV/u on four different targets: C, Al, Sn and Pb \cite{blank}. 
No experimental data are available here for momentum distributions. The
introduction of the final state interaction in the Coulomb dissociation part
does not change the result by much, compared with our previous analysis \cite%
{tra02}. We find now $C_{p_{3/2}}^{2}+C_{p_{1/2}}^{2}=1.26\pm 0.13$ fm$^{-1}$%
. To calculate the astrophysical $S$-factor we use the potential model. We
find $S_{18}(0)=46\pm 6$ eV$\cdot $b. A very weak dependence on energy is
observed: $S(E)=45.8-15.1E+7.34E^{2}$ ($E$ in MeV). This result is in very
good agreement with other determinations \cite{beaumel,enders}, but not with
one from Coulomb dissociation \cite{motob}, a fact that we do not understand. 
We underline that for this case
the precision achieved from this determination is the best so far and is
sufficient for astrophysical purposes.

\subsection{Breakup of $^{23}$Al and the consequences on the $^{22}$Mg(p,$%
\protect\gamma $)$^{23}$Al stellar reaction rate}

Space-based gamma-ray telescopes have the ability to detect $\gamma $-rays
of cosmic origin. They already provided strong and direct evidence that
nucleosynthesis is an ongoing process through the detection of transitions
in the decay of $^{26}$Al, $^{56}$Ni, $^{44}$Ti, etc. Among the expected $%
\gamma $-ray emitters is $^{22}$Na (T$_{1/2}$=2.6 y) produced in the
thermonuclear runaway and the high-temperature phase in the so-called ONe
novae (Oxygen-Neon novae) through the reaction chain $^{20}$Ne(p,$\gamma $)$%
^{21}$Na(p,$\gamma $)$^{22}$Mg($\beta $,$\gamma $)$^{22}$Na (NeNa cycle) 
\cite{starrfield98,jose99,wanajo99}. Measurements, however, have not 
detected the 1.275 MeV gamma-ray following the decay of $^{22}$Na and
have only been able to set an upper limit on its production, a limit which
is below the theoretical predictions (see, for example, \cite{iyudin95}
and the references therein). This discrepancy may arise from a poor
knowledge of the reaction cross sections employed in the network
calculations for the rp-process. In particular, it was proposed that the
precursor $^{22}$Mg can be depleted by the radiative proton capture reaction 
$^{22}$Mg(p,$\gamma $)$^{23}$Al \cite{wiescher88}, which can result in a
serious reduction of the $^{22}$Na abundance. The reaction is dominated by
direct capture and resonant capture through the first excited state
in $^{23}$Al. There is no direct measurement of the cross section at stellar
energies because it is impossible to make a $^{22}$Mg (T$_{1/2}$=3.86 s) 
target and difficult to obtain an intense $^{22}$Mg beam. Therefore, 
currently the rate of this reaction is estimated based on the mass
and resonance energy determined experimentally \cite{caggiano01} and
assuming that the spins and parities are as in the mirror system $^{23}$Ne.

\begin{figure}[tbp]
\resizebox{0.45\textwidth}{!}{     \includegraphics{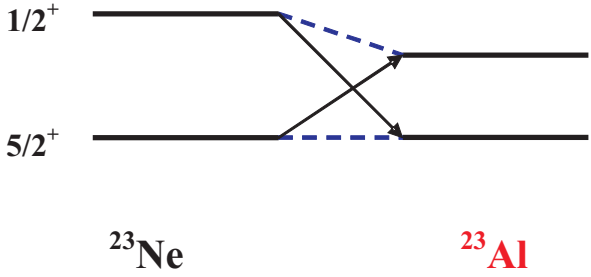}}
\caption{{\protect\footnotesize The level inversion in $^{23}$Al suggested
in Refs. {\protect\cite{cai02,zhang02}}.}}
\label{fig2}
\end{figure}

\begin{figure}[tbp]
\resizebox{0.45\textwidth}{!}{     \includegraphics{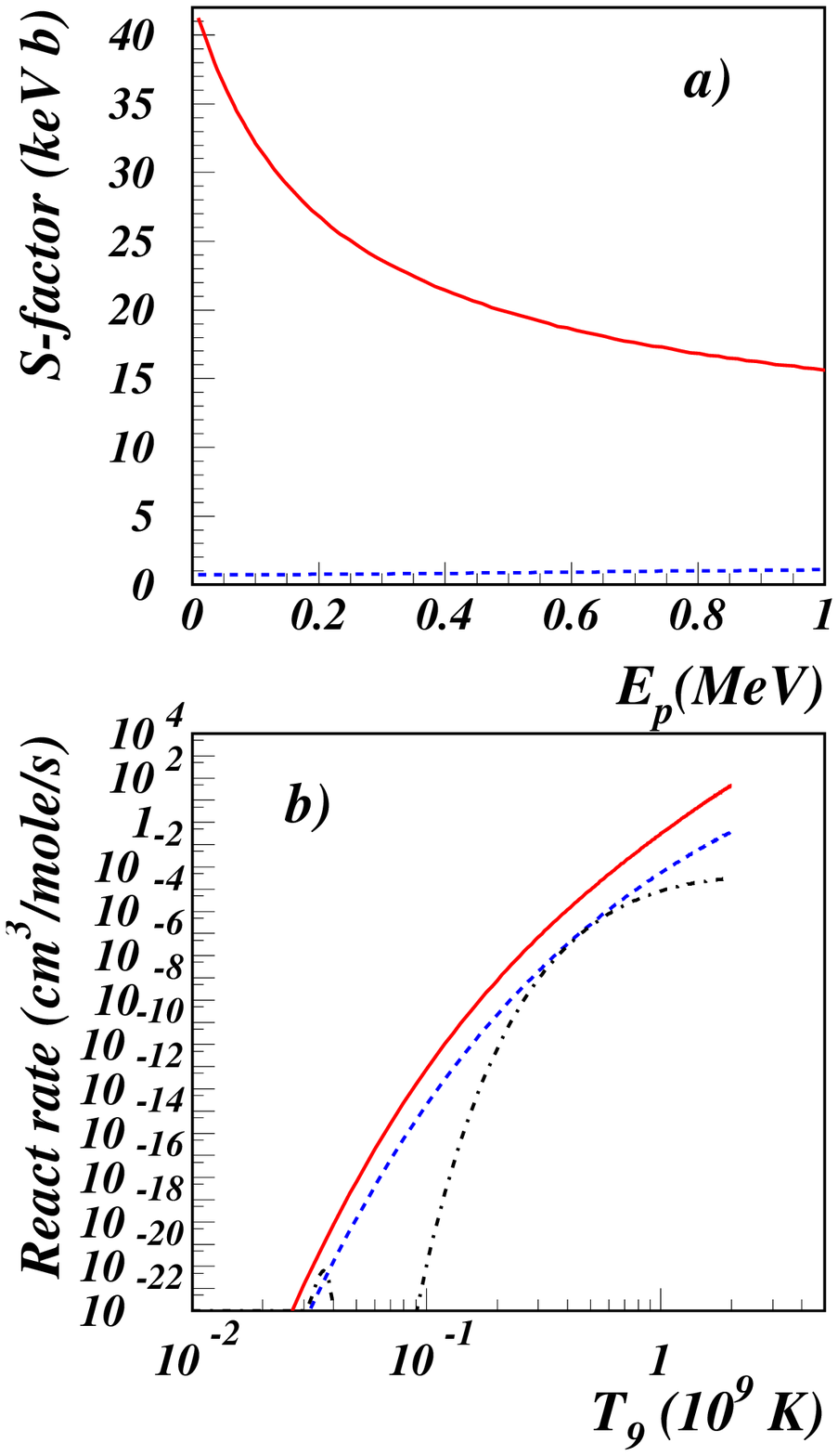}}
\caption{{\protect\footnotesize a) The astrophysical S-factor for the $^{22}$%
Mg(p,$\protect\gamma $)$^{23}$Al reaction, calculated assuming $J^{\protect%
\pi }$=$5/2^+$ (dashed line), or $J^{\protect\pi }$=$1/2^+$ (orbital
inversion, full line) for the g.s. of $^{23}$Al. b) The corresponding
reaction rate calculated for the case of non-inversion (dashed), or
inversion (full line). The dash-dot line shows the resonant contribution of
the 1-st excited state in $^{23}$Al.}}
\label{fig3}
\end{figure}

The nucleus $^{23}$Al is a weakly bound proton rich nucleus ($%
S_{p}=0.123(19) $ MeV) close to the drip line. Recent measurements of the
reaction cross sections for N=10 isotones and Z=13 isotopes around 30
MeV/nucleon on a $^{12}$C target found a remarkable enhancement for $^{23}$%
Al, which led the authors to the conclusion that it is one of the rare
proton halo nuclei \cite{cai02}. This is explained with a presumed level
inversion between the $2s_{1/2}$ and $1d_{5/2}$ orbitals (Figure \ref{fig2}).
The inversion was
further supported by several microscopic nuclear structure calculations that
find J$^{\pi }$=1/2$^{+}$ for the $^{23}$Al ground state \cite{zhang02}. If
the above mentioned inversion is correct, it will affect the radiative
capture cross section much more strongly than any other uncertainties.
Indeed, assuming such an inversion, we recalculate the astrophysical $S$%
-factor (Figure \ref{fig3}a) and the stellar reaction rate (Figure \ref{fig3}b)
for the $^{22}$%
Mg(p,$\gamma $)$^{23}$Al reaction and find an increase of 30 to 50 times
over the current estimate of the rate for the temperature range $%
T_{9}=0.1-0.3$. Clearly then, it is important to determine the spin and
parity of the low-lying levels in $^{23}$Al. It is important for both
nuclear structure and for its consequences for nuclear astrophysics. 
As a further complication, the NNDC data base gives J$^{\pi}$=3/2$^+$ for 
the ground state of $^{23}$Al \cite{nndc}. 

\begin{figure*}[tbp]
\resizebox{0.85\textwidth}{!}{  \includegraphics{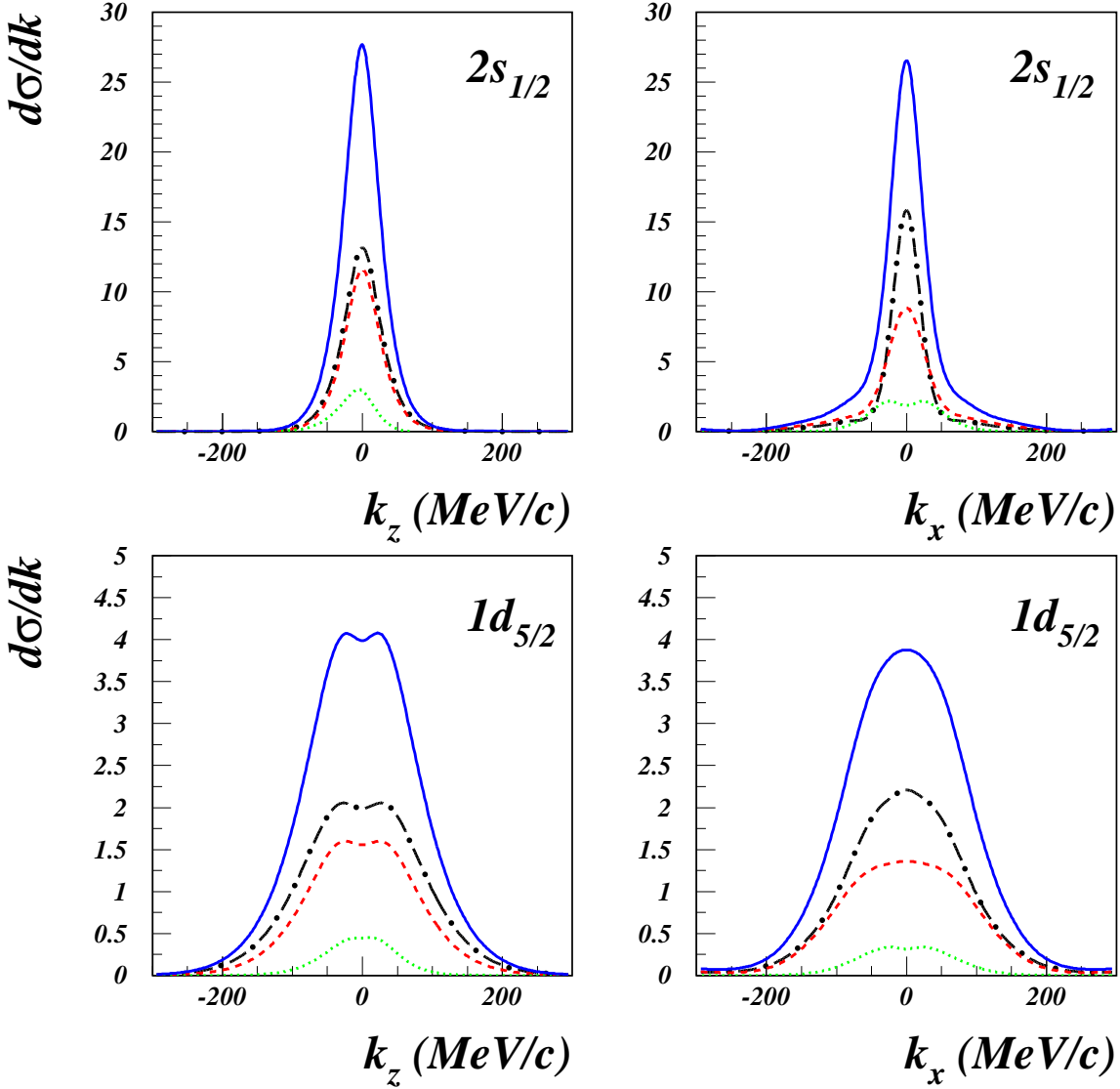}}
\caption{{\protect\footnotesize The calculated parallel (left panels) and 
transverse (right panels) momentum
distributions for the proton-removal (breakup) of 60 MeV/nucleon $^{23}$Al
on a light target in the two spin assumptions. The curves show the different
contributions: stripping (dash-dotted), diffraction dissociation (dashed)
and Coulomb (dotted). The full lines are the sum of all contributions. They
were all calculated assuming pure $1d_{5/2}$ or $2s_{1/2}$ orbitals,
respectively.}}
\label{fig4}
\end{figure*}

We proposed the use of intermediate-energy one-proton removal reactions on a
light target as a means to determine the structure of the $^{23}$Al ground
state. Such reactions have proven to be a reliable spectroscopic tool, with
advantages in particular for the case of weakly bound isotopes, close to the
drip lines \cite{hansen01,sauvan}. We calculate that for $^{12}$C($^{23}$Al,$%
^{22}$Mg) at 60 MeV/nucleon, the parallel momentum distribution is some 2
times narrower for a $2s_{1/2}$ orbital than for a $1d_{5/2}$ orbital 
(Figure \ref{fig4}) 
and the associated cross section is about a factor two larger. We intend
to compare the calculated momentum distributions and cross sections with the
experimental ones and determine the spin and parity of the $^{23}$Al ground
state. We shall derive the related ANCs and from them, the astrophysical 
$S-$factor. Calculations for the
momentum distributions have been performed with procedures similar to those
used previously \cite{tra01,sauvan}. The first step is the calculation of
the single-particle density in $^{22}$Mg using a spherical HF+BCS
calculation with the density energy functional of Beiner and Lombard. The
experimental proton separation energy in $^{23}$Al, $S_{p}=0.123\ $MeV, was
reproduced. There are two possibilities for the spin-parity of the ground
state: J$^{\pi }$=5/2$^{+}$ or 1/2$^{+}$. Glauber model calculations have
therefore been performed for each case, assuming pure $1d_{5/2}$, or $%
2s_{1/2}$ orbitals, in order to demonstrate the sensitivity of the
technique. Scattering functions defining the stripping and diffraction
transition operators were generated with double-folding potentials using the
JLM effective interaction, renormalized as above. Calculations were done for
breakup on a $^{12}$C target to minimize the Coulomb effects. 

We draw two conclusions from the calculations:

1. The two possible assignments may be resolved on the basis of the
inclusive cross sections and momentum distributions. For example the cross
section drops by a factor of two if a $1d_{5/2}$ state is assumed rather
than $2s_{1/2}$. This is easy to understand because the low-binding energy
and the lack of a centrifugal barrier in the case of a $2s_{1/2}$ orbital
leads to a much longer tail of the radial wave function than for the case of
the $1d_{5/2}$ orbital. Moreover, the very peripheral character of
single-nucleon removal reactions means that it is the asymptotic part of the
wave function that dictates the cross section and momentum distribution. In
the case of the latter, the widths of the momentum distributions differ by a
factor of two for both the parallel and transverse momenta, reflecting the
different behavior of the tails of the wave functions. Cross sections of 97
mb ($2s_{1/2}$), and of 42 mb ($1d_{5/2}$) were found for the two ground
state spin-parity assignments. The corresponding widths (FWHM) of the
distributions are predicted to be 60 MeV/c, and 180 MeV/c, respectively.

2. The shape of momentum distributions is extremely selective - narrow for a 
$2s_{1/2}$ state and broad with a flat top and a small central dip for
removal of a $1d_{5/2}$ (Figure 4). 

A study of one-proton removal from $^{23}$Al should, therefore, allow the
spin-parity of the ground state of $^{23}$Al to be deduced. Measurements of
the cross sections and momentum distributions in coincidence with gamma-rays
from the $^{22}$Mg core will allow us to disentangle the detailed structure
of the wave function, and in particular to deduce the spectroscopic factors
for the various configurations. This spectroscopic information will also be
valuable to determine if $^{23}$Al is deformed or spherical.

\section{Conclusions}

\label{sec:4}

In conclusion, we have shown that one-proton-removal reactions at intermediate
energies can be used to obtain astrophysical $S$-factors at stellar energies
for radiative proton capture reactions. Difficult or impossible direct
measurements for nuclear astrophysics at very low energies can be replaced
by indirect measurements with radioactive beams at larger energies. We find
that a kinematic window exists at 30-150 MeV/nucleon where the reactions are
peripheral and the relevant ANC can be determined. The method is
particularly useful because it can be used for rare isotopes, for poor
quality radioactive beams obtained from fragmentation, with cocktail beams
and with low intensity beams. It was shown that breakup at intermediate 
energies can be studied with beams as low as a few particles/sec 
\cite{maddalena01}. Our results from the use of
different NN interactions remind us of the fact that the precision of all
indirect methods depends not only on the precision of the experiments but
also on the accuracy of the calculations. Our findings may give a measure of
the present status of accuracy.

This work was supported in part by the U. S. Department of Energy under
Grant No. DE-FG03-93ER40773, by the Romanian Ministry for Research and
Education under contract no 555/2000, and by the Robert A. Welch Foundation.


\end{document}